# Between-trial heterogeneity in meta-analyses may be partially explained by reported design characteristics

KM Rhodes[1], RM Turner[1,2], J Savović[3,4], HE Jones[3], D Mawdsley[5], JPT Higgins[3]

[1]MRC Biostatistics Unit, School of Clinical Medicine, University of Cambridge, Cambridge, UK
[2] MRC Clinical Trials Unit, University College London, UK
[3]Population Health Sciences, Bristol Medical School, University of Bristol, UK
[4]NIHR CLAHRC West, University Hospitals Bristol NHS Foundation Trust, Bristol, UK
[5]University of Manchester, Manchester, UK

*Abstract*

**Objective**

We investigated the associations between risk of bias judgments from Cochrane reviews for sequence generation, allocation concealment and blinding and between-trial heterogeneity.

**Study Design and Setting**

Bayesian hierarchical models were fitted to binary data from 117 meta-analyses, to estimate the ratio $\lambda$ by which heterogeneity changes for trials at high/unclear risk of bias, compared to trials at low risk of bias. We estimated the proportion of between-trial heterogeneity in each meta-analysis that could be explained by the bias associated with specific design characteristics.

**Results**

Univariable analyses showed that heterogeneity variances were, on average, increased among trials at high/unclear risk of bias for sequence generation ($\hat{\lambda}$ 1.14, 95% interval: 0.57 to 2.30) and blinding ($\hat{\lambda}$ 1.74, 95% interval: 0.85 to 3.47). Trials at high/unclear risk of bias for allocation concealment were on average less heterogeneous ($\hat{\lambda}$ 0.75, 95% interval: 0.35 to 1.61). Multivariable analyses showed that a median of 37% (95% interval: 0% to 71%) heterogeneity variance could be explained by trials at high/unclear risk of bias for sequence generation, allocation concealment and/or blinding. All 95% intervals for changes in heterogeneity were wide and included the null of no difference.

**Conclusion**

Our interpretation of the results is limited by imprecise estimates. There is some indication that between-trial heterogeneity could be partially explained by reported design characteristics, and hence adjustment for bias could potentially improve accuracy of meta-analysis results.

# Introduction

In published meta-analyses, the original studies are often affected by varying amounts of internal bias caused by methodological flaws. Empirical studies have investigated the extent of between-study heterogeneity in a meta-analysis [1, 2]. This is likely to comprise a mixture of variation caused by true diversity among the study designs, variation due to within-study biases and unexplained variation. For this reason, it would be preferable to separate heterogeneity due to bias from other sources of between-study variation, as proposed by Higgins *et al.* [3].

Biases associated with reported study design characteristics can be investigated within meta-epidemiological studies that analyse a collection of meta-analyses. An early example is that of Schulz *et al.* [4], where the methodological quality of 250 randomized controlled trials from 33 meta-analyses within the Cochrane Pregnancy and Childbirth database was assessed. Schulz *et al.* provided empirical evidence to suggest that trials in which randomization is inadequately concealed report exaggerated estimates of intervention effect compared with adequately concealed trials. There was also some indication that trials with inadequate blinding yield larger effect estimates.

More recently, the *BRANDO* (Bias in Randomized and Observational Studies) study and *ROBES* (Risk of Bias in Evidence Synthesis) study have investigated the associations between reported design characteristics and intervention effects and heterogeneity [5, 6]. The *BRANDO* study combined data from all existing meta-epidemiological studies (collections of meta-analyses) into a single database, comprising 1973 independent trials included in 234 meta-analyses. The *ROBES* database included 228 binary outcome meta-analyses from Cochrane reviews that had implemented the Cochrane risk-of-bias tool [7]. In the *BRANDO* study, all trials included in the database had been categorised according to whether they were judged as adequate, inadequate, or unclear for sequence generation, allocation concealment and double-blinding. Trials included in the *ROBES* study database had been categorised as being at high, low or unclear risk of bias for sequence generation, allocation concealment, blinding and incomplete outcome data, using the Cochrane risk-of-bias tool. The results of both meta-epidemiological studies showed that the relative intervention effect in favour of the experimental treatment is, on average, modestly exaggerated in trials with inadequate randomization and lack of blinding. The *ROBES* study found no evidence of bias due to a high or unclear risk of bias assessment for incomplete outcome data. Both studies also found that bias in intervention effect estimates associated with the lack of blinding in trials with subjective outcome measures may be unpredictable in its direction and magnitude, leading to increased within meta-analysis heterogeneity.

When deciding how to handle suspected biases, meta-analysts often consider whether to restrict their analyses to studies at lower risk of bias or to include all available evidence. Restricting analyses to studies at lower risk of bias may lead to an unbiased result, but this result would be imprecise if high quality evidence is sparse. On the other hand, combining all available studies and ignoring flaws in their conduct could lead to biased summary estimates with inappropriate clinical or policy decisions as a possible consequence. Welton *et al.* [8] proposed a method for meta-analysis that uses all available data, while adjusting for and down-weighting the evidence from lower quality studies, based on evidence from a meta-epidemiological study.

The analyses of *BRANDO* and *ROBES* followed methods proposed by Welton *et al.* [8], which model the effects of lower quality design characteristics on average bias and between-trial heterogeneity. Under these models, the trials judged to be of poorer quality were assumed to be at least as heterogeneous as those of higher quality, which may not be the case. We have since proposed label-invariant models that avoid this constraint [9]. Our models are more flexible than the models of Welton *et al.* in allowing us to quantify the ratio by which heterogeneity changes for studies with lower quality design characteristics. This facilitates investigation of how much between-study heterogeneity in a meta-analysis is attributable to lower quality studies.

Bias can lead to overestimation or underestimation of the true intervention effect in a study, and we could expect differences in risk of bias across studies to contribute to variation among the results of studies included in a meta-analysis. Here we re-analyse trial data from the *ROBES* database, using our label-invariant models to investigate the associations between risk of bias judgments from Cochrane reviews and heterogeneity among randomized controlled trials. We investigate the extent of heterogeneity in a meta-analysis that is due to within-trial biases. The empirical evidence provided gives useful information on the extent to which we might expect the between-trial variance to change in a meta-analysis, if we adjust for known sources of bias.

**Methods**

**Data description**
We make use of data from the *ROBES* (Risk of Bias in Evidence Synthesis) [5] study, which is a large collection of meta-analyses extracted from the Cochrane Database of Systematic Reviews. These data were originally used to examine the associations between reported design characteristics and intervention effect estimates in meta-analyses. Meta-analyses with fewer than five trials were excluded, as were meta-analyses where the review authors considered pooling to be inappropriate or where numerical data were unavailable. One or more binary outcome meta-analysis from each eligible review was included in the database, corresponding to a primary outcome where possible.

The dataset includes 228 meta-analyses from Cochrane reviews that had information on all five of the following Risk of Bias items: sequence generation; allocation concealment; blinding; incomplete outcome data and selective outcome reporting. In this paper we do not consider the influence of accounting for bias caused by incomplete outcome data or selective outcome reporting on heterogeneity. The *ROBES* study found no evidence of exaggerated intervention effect among trials at high or unclear risk of bias (compared with low risk of bias) for assessment of incomplete outcome data [5] and it is not generally recommended to try to adjust for selective outcome reporting bias in meta-analysis [10].

Our statistical analyses were carried out on a subset of the *ROBES* study, comprising 1473 trials from 117 meta-analyses. These meta-analyses contained at least one trial at low risk of bias and at least one trial at high or unclear risk of bias for each of the three characteristics of interest: sequence generation, allocation concealment and blinding. Focusing on one subset of the data throughout all analyses allowed for direct comparison of results assessing the influences of accounting for different combinations of study design characteristics on heterogeneity. Table 1 shows the structure of the dataset. For each trial included in the *ROBES* database, we have binary outcome data consisting of the number of events in each treatment arm and the total number of participants in each arm. The direction of outcome events in the *ROBES* database is coded such that the outcome for each trial corresponds to a harmful event. All meta-analyses in the database have been categorised according to the type of outcome under assessment and the types of interventions evaluated, in the same way as Turner *et al*. [1]. Outcomes in the *ROBES* database were classified into three broad categories (all-cause mortality, other objective, subjective) in the same way as the *BRANDO* study [6].

**Table 1** Structure of the dataset

|  | *N* | Min | Median | Max | IQR |
|---|---|---|---|---|---|
| **No. of trials per meta-analysis** | 117 meta-analyses | 5 | 10 | 75 | 6 to 14 |
| **No. of participants per trial** | 1473 trials | 8 | 119 | 182,000 | 60 to 267 |

**Statistical analysis**

We used label-invariant hierarchical models to analyse trial data from all included meta-analyses simultaneously. The models were fitted as described in an earlier paper [9] and are based on an extension of the model described as "Model 3" by Welton *et al*. [8]. Within each meta-analysis, a model with binomial within-trial likelihoods was fitted to the binary outcome data from each trial on the log odds ratio scale. The model assumes that the higher quality trials at low risk of bias provide an unbiased estimate of intervention effect, assumed to have a normal random-effects distribution with variance $\tau_m^2$ specific to each meta-analysis indexed *m*. Throughout our analyses, we used a dichotomised variable for each design characteristic (high or unclear risk of bias compared with low risk of bias for sequence generation, allocation concealment and blinding). The trials at high or unclear risk of bias are assumed to estimate the sum of two components: the same intervention effect as the trials at low risk of bias plus some trial-specific bias. Within each meta-analysis, we quantify variation among trials at high or unclear risk of bias by $\lambda\tau_m^2$, which can be lower or higher than the variation $\tau_m^2$ among trials at low risk of bias. For each design characteristic, the hierarchical models allow us to estimate: the average bias in estimated intervention effect within meta-analysis *m* ($b_m$); the average bias in estimated intervention effect across meta-analyses ($b_0$); the ratio by which between-trial heterogeneity in intervention effects changes for trials with potential flaws ($\lambda$); and variation in average bias across meta-analyses ($\varphi$).

We first conducted univariable analyses examining the influence of accounting for a single trial design characteristic on heterogeneity before carrying out multivariable analyses examining the influence of accounting for all three design characteristics. In multivariable analyses, interactions between the different design characteristics were assumed to have distinct variance components $\lambda$ and $\varphi$.

Following the approach of Turner *et al.* [1], we fitted a log-normal model to underlying values of heterogeneity variance $\tau_m^2$ in intervention effect among trials with low risk of bias across meta-analyses. Previous research has shown that the extent of total heterogeneity in a meta-analysis differs according to the type of outcome examined in the meta-analysis [1, 2]. To investigate association between the type of outcome under assessment and the heterogeneity variance among trials with low risk of bias, we included indicators for the different types of outcome as covariates in the model for $\tau_m^2$.

All models were fitted using Markov chain Monte Carlo (MCMC) methods within *WinBUGS* Version 1.4.3 [11]. We based results on 100,000 iterations, following a burn-in period of 10,000 iterations, which was sufficient to achieve convergence and produced low MC error rates. Convergence was assessed according to the Brooks-Gelman-Rubin diagnostic tool [12], using two chains starting from widely dispersed initial values. As in our earlier paper [9], we assigned normal(0,1000) prior distributions to location parameters and a log-normal(0,1) prior to $\lambda$. Variation in average bias across meta-analyses, $\varphi$, was assigned an inverse-gamma(0.001,0.001) prior with increased weight on small values. Model fit was assessed using the deviance information criterion (DIC), as recommended by

Spiegelhalter *et al.* [13,14]. Due to the non-linearity between the likelihood and the model parameters, we calculated the effective number of parameters at the posterior mean of the fitted values rather than at the posterior mean of the basic model parameters [15]. The *WinBUGS* code for fitting the label-invariant models is available in the Supporting Information of an earlier paper [9].

**Quantifying heterogeneity due to bias**

It is of interest to quantify the proportion of between-trial heterogeneity in a meta-analysis that can be explained by the bias associated with reported design characteristics. This requires an estimate of total heterogeneity variance among all trials included in a meta-analysis and an estimate of the heterogeneity variance after accounting for biases. The latter is estimated from the model above, where the three design characteristics are assumed to be responsible for all of the within-trial biases. In univariable analyses for the influence of accounting for a single characteristic, we estimated total heterogeneity variance $\tau^2_{total,m}$ in a meta-analysis $m$, using the formula $\tau^2_m / ((1-\pi_m)\tau^2_m + \pi_m \lambda \tau^2_m + \pi_m(1-\pi_m)b^2_m)$, where $\pi_m$ is the proportion of trials at high or unclear risk of bias in meta-analysis $m$. The derivation of this formula for total heterogeneity variance is provided in Supplementary material (S1), together with the formula used in multivariable analyses for the influence of accounting for two characteristics. We note that the formula used in multivariable analyses for the influence of accounting for three characteristics is derived in the same way. For each meta-analysis $m$ within the subset of 117 meta-analyses in *ROBES*, we used *WinBUGS* to obtain the posterior median for the ratio of between-trial variance among trials at high or unclear risk of bias to total between-trial variance $1 - \tau^2_m / \tau^2_{total,m}$. For each individual design characteristic and all combinations of design characteristics, we summarise the proportion of heterogeneity attributable to trials at high or unclear risk of bias by the median and 95% interval of posterior medians for $1 - \tau^2_m / \tau^2_{total,m}$ across meta-analyses indexed $m$.

Negative estimates of the proportion of heterogeneity due to trials at high or unclear risk of bias occur where the estimate of total heterogeneity variance $\tau^2_{total}$ among all trials included in a meta-analysis is less than the estimate of the heterogeneity variance $\tau^2$ among trials at low risk of bias. We note that $\tau^2_{total}$ is not only increased from $\tau^2$ by the heterogeneity variance among trials at high or unclear risk of bias, but also the difference in intervention effect between the trials at high or unclear risk of bias and the trials at low risk of bias (see formula in Supplementary material (S1)). We set the negative values of the ratio to zero, since total between-trial heterogeneity in the meta-analysis cannot be explained by the trials at high or unclear risk of bias.

We graphically explored the influence of accounting for reported design characterises on heterogeneity on randomized trials in meta-analysis. For each meta-analysis $m$ within the subset of *ROBES*, we plotted the posterior median of heterogeneity variance $\tau_m^2$ among trials at low risk of bias against the posterior median of heterogeneity variance $\tau_{total,m}^2$ among all trials.

**Results**

**Descriptive analyses**

Table 2 reports the number of trials with each combination of reported design characteristics. The frequency of trials categorised as being at high or unclear risk of bias for a single design characteristic was 303 (21%), of which 75 (25%) were at high or unclear risk of bias for sequence generation, 98 (32%) were at high or unclear risk of bias for allocation concealment and 130 (43%) were at high or unclear risk of bias for blinding. The number of trials categorised as being at high or unclear risk of bias for precisely two design characteristics was somewhat higher at 413 (28%). All three design characteristics were judged as high or unclear risk in 396 (27%) of trials. For each design characteristic, Table 2 shows the breakdown of the trial numbers into high risk of bias and unclear risk of bias, overall and according to the type of outcome under assessment. Of all 1473 trials in the dataset, sequence generation was assessed as high risk of bias in 41 (3%) trials, unclear in 736 (50%) trials, and low risk of bias in 696 (47%) trials. Allocation concealment was assessed as high risk of bias in 80 (5%) trials, unclear in 760 (52%) trials, and low risk of bias in 633 (43%) trials. Blinding was assessed as high risk of bias in 317 (22%) trials, unclear in 383 (26%) trials, and low risk of bias in 773 (52%) trials. The proportions of trials judged as being at high or unclear risk of bias are greatest among trials with subjectively measured outcomes, and lowest among trials assessing all-cause mortality.

**Table 2** The overall number of trials with each combination of reported design characteristics, within the subset of 117 meta-analyses extracted from *ROBES*, and the number of trials at high or unclear risk of bias for each reported design characteristics overall and according to type of outcome measure.

| Risk of bias | | | No. of trials (%) | No. of trials at high risk of bias (% of trials) | | | No. of trials at unclear risk of bias (% of trials) | | |
|---|---|---|---|---|---|---|---|---|---|
| Sequence generation | Allocation concealment | Blinding | | Sequence generation | Allocation concealment | Blinding | Sequence generation | Allocation concealment | Blinding |
| Low | Low | Low | 361 (25%) | - | - | - | - | - | - |
| High or unclear | Low | Low | 75 (5%) | 0 | - | - | 75 (100%) | - | - |
| Low | High or unclear | Low | 98 (7%) | - | 8 (8%) | - | - | 90 (92%) | - |
| Low | Low | High or unclear | 130 (9%) | - | - | 75 (58%) | - | - | 55 (42%) |
| High or unclear | High or unclear | Low | 239 (16%) | 9 (4%) | 8 (3%) | - | 230 (96%) | 231 (97%) | - |
| High or unclear | Low | High or unclear | 67 (5%) | 1 (1%) | - | 28 (42%) | 66 (99%) | - | 39 (58%) |
| Low | High or unclear | High or unclear | 107 (7%) | - | 19 (18%) | 60 (56%) | - | 88 (82%) | 47 (44%) |
| High or unclear | High or unclear | High or unclear | 396 (27%) | 31 (8%) | 45 (11%) | 154 (39%) | 365 (92%) | 351 (89%) | 242 (61%) |
| | | Overall | 1473 (100%) | 41 (3%) | 80 (5%) | 317 (22%) | 736 (50%) | 760 (42%) | 383 (26%) |
| | | Mortality outcome | 271 (18%) | 7 (3%) | 22 (8%) | 76 (28%) | 100 (37%) | 104 (38%) | 36 (13%) |
| | | Objective outcome[1] | 301 (20%) | 9 (3%) | 13 (4%) | 74 (25%) | 145 (48%) | 152 (51%) | 54 (18%) |
| | | Subjective outcome[2] | 901 (61%) | 25 (3%) | 45 (5%) | 167 (19%) | 491 (55%) | 504 (56%) | 293 (33%) |

[1] 10 (37%) meta-analyses measured objective outcomes other than all-cause mortality including laboratory assessed outcomes, pregnancy and perinatal outcomes. 17 (62%) meta-analyses assessed objective outcomes potentially influenced by judgment such as caesarean section and hospital admissions; [2] Subjectively measured outcomes include pain, mental health outcomes, cause-specific mortality, clinically-assessed outcomes, signs and symptoms reflecting continuation/end of condition and lifestyle outcomes.

**Model comparison**

Results from model comparison are provided in Supplementary material (S2). The multivariable model for the influence of accounting for high or unclear risk of bias for sequence generation and blinding (Model B2) had an improved fit when an interaction term was included. However, after adjustment for trials at high or unclear risk of bias for allocation concealment (Model B4) there was no evidence of interaction between sequence generation and blinding. Despite this, we base our results on models including interaction terms among reported design characteristics, because we would expect reported design characteristics to interact in practice.

The inclusion of outcome type indicators in the model for heterogeneity variance $\tau^2$ did not lead to a substantial improvement in model fit. For this reason, our results are based on hierarchical models for $\tau^2$ fitted without these covariates.

**Exploring the associations between reported trial design characteristics and heterogeneity**

Reported in Table 3 are estimates of $\lambda$ representing the ratio by which heterogeneity variance changes for trials at high or unclear risk of bias for specific design characteristics, compared to trials at low risk of bias. Estimates of average bias ($b_0$) and variation in mean bias across meta-analyses ($\varphi$) were almost identical to those reported elsewhere [5], and hence not reported here.

Each estimate of $\lambda$ in Table 3 is very imprecisely estimated; the 95% credible intervals for $\lambda$ are wide and contain the null value 1 representing no difference in heterogeneity among trials at high or unclear risk of bias and trials at low risk of bias. For this reason we interpret the results that follow with caution.

*Univariable analyses*

Based on univariable analyses for the influence of accounting for a single reported design characteristic, variation among trials at high or unclear risk of bias for sequence generation is, on average, 14% greater than that among trials at low risk of bias for sequence generation ($\hat{\lambda}$ 1.14, 95% interval: 0.57 to 2.30). Heterogeneity among trials judged as high or unclear risk of bias for allocation concealment is, on average, 75% that among trials assessed as low risk of bias for allocation concealment ($\hat{\lambda}$ 0.75, 95% interval: 0.35 to 1.61). The central estimate for $\lambda$ suggests that variation among trials at high or unclear risk of bias for blinding is, on average, 74% greater than that among trials at low risk of bias for blinding ($\hat{\lambda}$ 1.74, 95% interval: 0.85 to 3.47).

**Table 3** Results from univariable and multivariable analyses for the influence of accounting for trials at high or unclear risk of bias for specific design characteristics on heterogeneity. Posterior medians and 95% intervals are reported.

| Model | Univariable analyses | $\lambda$ |
|---|---|---|
| | High or unclear risk (vs low risk) of bias for: | |
| A1 | sequence generation | 1.14 (0.57 to 2.30) |
| A2 | allocation concealment | 0.75 (0.35 to 1.61) |
| A3 | blinding | 1.74 (0.85 to 3.47) |
| | **Multivariable analyses (from models including interaction terms)*** | |
| | High or unclear risk (vs low risk) of bias for: | |
| B1 | sequence generation, in trials at low risk of bias for allocation concealment | 0.76 (0.14 to 1.79) |
| | allocation concealment, in trials at low risk of bias for sequence generation | 0.54 (0.10 to 1.41) |
| | sequence generation and allocation concealment | 0.94 (0.39 to 1.90) |
| B2 | sequence generation, in trials at low risk of bias for blinding | 0.59 (0.14 to 1.46) |
| | blinding, in trials at low risk of bias for sequence generation | 1.01 (0.41 to 2.73) |
| | sequence generation and blinding | 1.58 (0.59 to 4.65) |
| B3 | allocation concealment, in trials at low risk of bias for blinding | 0.65 (0.20 to 2.14) |
| | blinding, in trials at low risk of bias for allocation concealment | 1.69 (0.44 to 5.68) |
| | allocation concealment and blinding | 1.41 (0.55 to 4.02) |
| B4 | sequence generation, in trials at low risk of bias for allocation concealment & blinding | 0.46 (0.11 to 1.13) |
| | allocation concealment in trials at low risk of bias for sequence generation & blinding | 0.49 (0.12 to 1.71) |
| | blinding, in trials at low risk of bias for sequence generation & allocation concealment | 0.99 (0.43 to 2.31) |
| | sequence generation and allocation concealment, in trials at low risk of bias for blinding | 0.39 (0.07 to 1.29) |
| | sequence generation and blinding, in trials at low risk of bias for allocation concealment | 1.44 (0.34 to 5.34) |
| | allocation concealment and blinding, in trials at low risk of bias for sequence generation | 0.50 (0.16 to 1.92) |
| | sequence generation, allocation concealment and blinding | 1.22 (0.39 to 3.01) |

$\lambda$ ratio of heterogeneity variance among trials at high or unclear risk of bias to heterogeneity variance among trials at low risk of bias.

*Note that results for multiple characteristics are not implied by the results for each individual bias domain in the multivariable analysis, due to the presence of all possible interactions between bias domains.

*Multivariable analyses*

Also reported in Table 3 are results from multivariable analyses for the influence of accounting for combinations of design characteristics. Based on results from fitting Model B1, heterogeneity among trials at high or unclear risk of bias for both sequence generation and allocation concealment is, on average, 94% that among trials at low risk of bias for both sequence generation and allocation concealment ($\hat{\lambda}$ 0.94, 95% interval: 0.39 to 1.90). Heterogeneity among trials at high or unclear risk of bias for both sequence generation and blinding is, on average, 58% greater than that among trials at low risk of bias for both characteristics based on results from fitting Model B2 ($\hat{\lambda}$ 1.58, 95% interval: 0.59 to 4.65). Results from fitting Model B3 show that heterogeneity is, on average, 41% greater among trials at high or unclear risk of bias for both allocation concealment and blinding, compared with trials at low risk of bias for both characteristics ($\hat{\lambda}$ 1.41, 95% interval: 0.55 to 4.02). Results from multivariable analyses for the influence of accounting for all three design characteristics (Model B4) imply that heterogeneity is, on average, 22% greater among trials at high or unclear risk of bias (compared with low risk of bias) for all three reported design characteristics ($\hat{\lambda}$ 1.22, 95% interval: 0.39 to 3.01). As in univariable analyses, estimates of association between heterogeneity and reported design characteristics are very uncertain; 95% credible intervals for $\lambda$ all contain the null effect.

**Investigating the extent of heterogeneity due to reported trial design characteristics**

We investigate the extent to which one might expect between-trial heterogeneity in a random-effects meta-analysis to change, on average, if we adjust for potential bias attributable to specific design characteristics in a new meta-analysis.

Table 4 summarises posterior medians of the proportion of total between-trial heterogeneity attributable to trials at high or unclear risk of bias across the subset of 117 meta-analyses in *ROBES*.

*Univariable analyses*

In univariable analyses for the influence of accounting for a single reported design characteristic, central estimates for the proportion of between-trial variance explained by trials at high or unclear risk of bias for sequence generation have median 30% (95% interval: 7% to 46%) across meta-analyses (Model A1). There is less evidence that between-trial heterogeneity in a meta-analysis is attributable to the bias associated with low or unclear quality for allocation concealment; central estimates for the proportion of heterogeneity among trials at high or unclear risk of bias have median 6% (95% interval: 0% to 17%) across meta-analyses (Model A2). Across meta-analyses, central estimates for the proportions of between-trial heterogeneity explained by bias associated with trials at high or unclear risk of bias for blinding have median 40% (95% interval: 8% to 56%) based on fitting Model A3.

For each of the 117 meta-analyses included within the subset of *ROBES*, Figure 1 presents a comparison of the central estimate of heterogeneity variance among trials at low risk of bias and the central estimate of heterogeneity variance among all trials. In separate univariable analyses for the influences of high or unclear risk of bias for sequence generation and blinding, the central estimate of heterogeneity variance among trials at low risk of bias tends to be lower than the central estimate of heterogeneity among all trials. In contrast, the central estimate of heterogeneity variance among trials at low risk of bias for allocation concealment is slightly higher than that among all trials in 73 (62%) meta-analyses.

*Multivariable analyses*

Based on results from multivariable analyses for the influence of accounting for multiple reported design characteristics, one might hypothesize that heterogeneity among trials in meta-analyses within *ROBES* can be explained by the bias associated with sequence generation and/or allocation concealment (Model B1); across meta-analyses within the subset of *ROBES,* central estimates for the proportion of heterogeneity due to trials at high or unclear risk of bias have median 19% (95% interval: 0% to 48%). Estimates of the proportion of heterogeneity due to trials at high or unclear risk of bias due to sequence generation and/or blinding have median 37% (95% interval: 0% to 57%) across meta-analyses (Model B2). This median is slightly lower at 31% (95% interval: 0% to 51%) for heterogeneity variance explained by bias associated with trials at high or unclear risk of bias for allocation concealment and/or blinding (Model B3). Across meta-analyses in *ROBES*, central estimates for the proportion of between-trial heterogeneity explained by bias associated with trials at high or unclear risk of bias for sequence generation, allocation concealment and/or blinding have median 37% (95% interval: 0% to 71%) based on fitting Model B4.

In multivariable analyses for the influence of accounting for all three characteristics, the central estimate of heterogeneity variance among trials at low risk of bias for all three characteristics is lower than the central estimate of heterogeneity variance among all trials in the majority of 107 (91%) meta-analyses (Figure 1).

**Table 4** Summaries of posterior medians for the proportion of heterogeneity due to trials at high or unclear risk of bias for each design characteristic and combinations of design characteristics within the subset of 117 meta-analyses extracted from *ROBES*.

| Model | Design characteristic/s | Proportion of heterogeneity due to trials at high or unclear risk of bias for the design characteristic/s * |
|---|---|---|
| A1 | Sequence generation | Median 0.30; 95% interval 0.07 to 0.46 |
| A2 | Allocation concealment | Median 0.06; 95% interval 0 to 0.17 |
| A3 | Blinding | Median 0.40; 95% interval 0.08 to 0.56 |
| B1 | Sequence generation and/or allocation concealment | Median 0.19; 95% interval 0 to 0.48 |
| B2 | Sequence generation and/or blinding | Median 0.37; 95% interval 0 to 0.57 |
| B3 | Allocation concealment and/or blinding | Median 0.31; 95% interval 0 to 0.51 |
| B4 | Sequence generation, allocation concealment and/or blinding | Median 0.37; 95% interval 0 to 0.71 |

* Negative estimates suggest that heterogeneity among trials in a meta-analysis cannot be explained by trials at high or unclear risk of bias and were hence set to zero.

**Figure 1** For each of the 117 meta-analyses within the subset of *ROBES*, the central estimate of heterogeneity variance among trials at low risk of bias plotted against the central estimate of heterogeneity variance among all trials. Central estimates of heterogeneity variance are based on results from univariable model A1 for sequence generation, univariable model A2 for allocation concealment, univariable model A3 for blinding, and multivariable model B4 for sequence generation, allocation concealment and blinding. Solid lines indicate that estimates are identical.

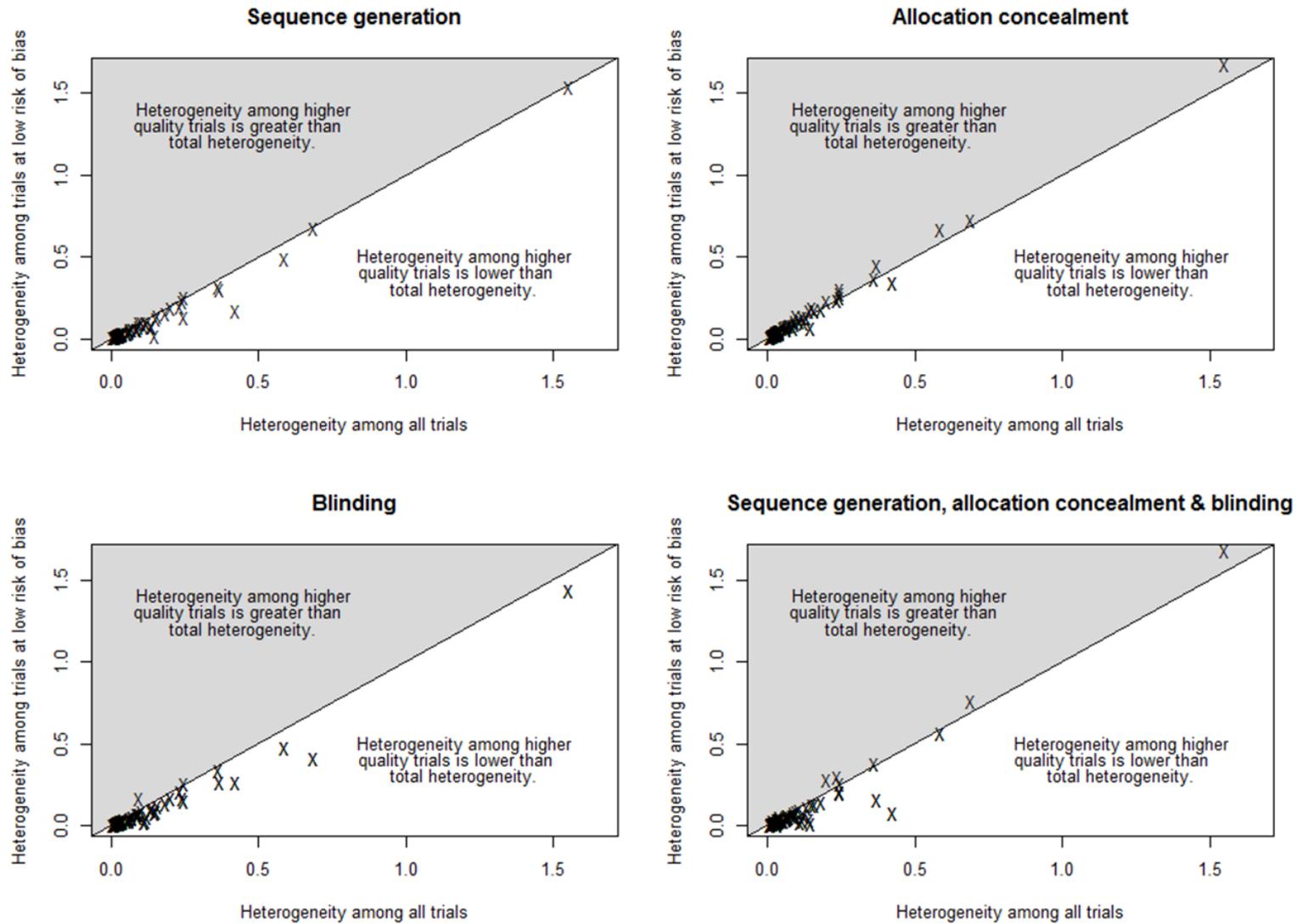

**Discussion**

Within-study biases can lead to overestimation or underestimation of the true intervention effect in a study and are expected to contribute to between-study variation in meta-analyses [5, 6, 16]. With access to a meta-epidemiological data set including meta-analyses which have implemented the Cochrane risk-of-bias tool, it was possible to explore the extent to which accounting for suspected biases influences levels of heterogeneity. We have investigated the impact of risk of bias judgments from Cochrane reviews for sequence generation, allocation concealment and blinding on between-trial heterogeneity, using data from 117 meta-analyses included in the *ROBES* study. Between-trial heterogeneity in intervention effect is a common problem in meta-analysis. The results of this empirical study show that roughly a third of between-trial heterogeneity might be explained by trial design characteristics, on average. Prediction intervals are becoming increasingly widely used to provide a predicted range for the true intervention effect in an individual study [3, 17], and are useful in decision making [18]. The implications of our research are that prediction intervals for true effects could be narrowed to account for biases, if they are to represent genuine variation in true effects.

This empirical study builds on previous meta-epidemiological studies [4-6] that have focussed on the influence of accounting for reported design characteristics on intervention effect rather than between-trial heterogeneity. Recent meta-epidemiological studies have tended to use the methods proposed by Welton *et al.*[8], which are less general in that they constrain trials at high or unclear risk of bias to be at least as heterogeneous as trials at low risk of bias. We previously proposed a more general model for the analysis of meta-epidemiological data [9]. In this study, the advantage of using our model was that we could estimate the quantity $\lambda$, representing the ratio by which heterogeneity changes for trials at high or unclear risk of bias, compared to trials at low risk of bias.

Random-effects meta-analysis may be appropriate when between-study heterogeneity exists. However, in some situations, studies differ substantially in quality so the random-effects assumption may be inadequate. When confronted with evidence of varying quality in practice, meta-analysts may decide to restrict their analyses to studies at lower risk of bias. However, this would not be practical in the typical situation where few studies are available to be included in the meta-analysis. The results of our meta-epidemiological study give some indication of increased heterogeneity among studies with high or unclear risk of bias judgements. These findings support recommendations to adjust for bias in meta-analyses of evidence of varying quality. Methods are available to adjust for and down-weight studies of lower quality in meta-analysis, using generic data-based evidence or expert opinion informed by detailed trial assessment [8, 19]. Based on our findings, these methods could be expected to reduce between-study variation in meta-analyses. Since the between-study variance parameter would be imprecisely estimated in many meta-analyses that only contain a small number of studies, we recommend assigning an informative prior distribution to this parameter, based on empirical evidence from historical meta-analyses [1, 2].

For each reported design characteristic and combinations of design characteristics, we calculated the proportion of heterogeneity in each meta-analysis that could be explained by trials at high or unclear risk of bias. Summaries of posterior medians for these proportions across meta-analyses give some indication of the reduction in between-trial heterogeneity we might expect to see in a meta-analysis, if we adjust for the bias associated with each reported design characteristic or combination of reported design characteristics. There is empirical evidence to suggest that flaws in the random sequence generation and lack of blinding may lead to increased levels of heterogeneity among randomized controlled trials, on average, but flawed methods of allocation concealment might have little impact. These findings should be interpreted with caution due to the limited statistical power to detect differences in heterogeneity between higher and lower quality trials. In each analysis the ratio of heterogeneity variance $\lambda$ attributable to bias was very imprecisely estimated. Although it would be expected for $\lambda$ to be imprecisely estimated in a single meta-analysis, we hoped to gain precision when estimating across the collection of meta-analyses included in the *ROBES* database; however, variability across meta-analyses was high.

In our analyses of the *ROBES* data, we wanted to allow the data to dominate and used a vague log-normal(0,1) prior distribution for the heterogeneity parameter $\lambda$. However, given the small amount of information available on $\lambda$ in the dataset, there was a possibility that results could have been sensitive to the choice of vague prior distribution. In an earlier paper, we used the same dataset in a sensitivity analysis to compare the effects of 5 different prior distributions for $\lambda$ [9]. Posterior estimates for the scale parameter $\lambda$ were consistent among the different priors, with similar medians and overlapping credible intervals.

Heterogeneity among trials at low risk of bias could be explained by clinical differences, for example difference in participants, or in the dosage or timing of an intervention. In each univariable and multivariable analysis, we did not find evidence of association between heterogeneity variance among trials at low risk of bias and the type of outcome under assessment in the meta-analysis. This might be explained by the fact that the majority of the outcomes examined in the meta-analyses included in our analyses were subjectively measured. In future work it would be of interest to explore how the extent of between-trial heterogeneity due to bias may depend on the type of outcome under assessment and the types of interventions being compared.

Another limitation is the accuracy of reported design characteristics which may not well represent how a trial was actually conducted. Trials that are conducted well could be poorly reported [20]. Hill

*et al.* [21] investigated discrepancies between published reports and actual conduct of randomized clinical trials and found that sequence generation and allocation concealment were reported as unclear in over 75% of studies where these two characteristics were actually at low risk of bias. A more recent study found that descriptions of blinding in trial protocols and corresponding reports were often in agreement [22]. These investigations provide some insight as to why the influences of accounting for high or unclear risk of bias for sequence generation and high or unclear risk of bias for allocation concealment on intervention effect and between-trial variance are smaller, compared with the effects of high or unclear risk of bias for blinding. Ideally we would have investigated this further, by separating trials at unclear risk of bias from trials at high risk of bias and comparing heterogeneity estimates between trials at high risk of bias and trials at low or unclear risk of bias. However the data on trials at high risk of bias were sparse.

It is possible that our results were confounded by the influence of other types of biases that could not be accounted for in our analyses. For example, there is empirical evidence of bias in the results of meta-analyses due to publication bias and selective reporting of outcomes arising from the lack of inclusion of statistically non-significant results [23, 24]. Methods to adjust for reporting biases are available, but it would have been impractical to apply these methods to each meta-analysis in our dataset. Meta-analyses affected by reporting biases would be expected to overestimate intervention effect and so the extent of heterogeneity that we observed among trials in the *ROBES* database could be higher than expected.

The *ROBES* dataset was extracted from the April 2011 issue of the *Cochrane Database of Systematic Reviews*, for which the risk of bias in trials may have been assessed prior to 2011. As of early 2011, Cochrane review authors have assessed risk of bias due to blinding of participants and personnel separately from blinding of outcome assessors. In the future, it would be of interest to investigate separate influences of accounting for blinding of participants and personnel and blinding of outcome assessors on intervention effect and between-trial heterogeneity, once large collections of meta-analyses with such assessments become available. It would also be of interest to investigate the impact of bias on intervention effect and heterogeneity in other types of meta-analyses; our analyses were conducted using binary outcome data from Cochrane reviews only. These include a wide range of application areas but may not be representative of all healthcare meta-analyses, and so the findings in this paper may not be generalizable to meta-analyses included in other systematic reviews.

In conclusion, the overall implications of this research are that the accuracy of meta-analysis results could be improved by adjusting for reported study design characteristics in the meta-analysis model. After conducting a random-effects meta-analysis, it is important to consider the potential effect of the intervention when it is applied within an individual study setting because this might be different from the average effect. In the presence of substantial heterogeneity among studies, prediction intervals for the true intervention effect in an individual study will be wide and uncertain. This empirical study gives some indication that adjustment for bias could reduce the uncertainty in predictive inferences, and better reflect the potential effectiveness of the intervention. A strategy of including all studies with such adjustments may produce a more favourable trade-off between bias and precision than excluding studies assessed to be at high risk of bias. However, interpretation of our results is limited by extremely imprecise estimates.

**Acknowledgements**

This project was supported by UK Medical Research Council (MRC) fellowship (G0701659/1) and grant (MR/K014587/1). KR and RT were supported by the MRC programme grant (U105260558), and RT also by the MRC grant MC_UU_12023/21, and JS was supported by the National Institute for Health Research (NIHR) Collaboration for Leadership in Applied Health Research and Care West (CLAHRC West) at University Hospitals Bristol NHS Foundation Trust. HEJ was supported by a MRC career development award in biostatistics (MR/M014533/1). The views expressed are those of the authors and not necessarily those of the MRC, the National Health Service, the NIHR or the Department of Health.

# Supplementary materials

## S1 Estimating total heterogeneity variance from the label-invariant model

We used label-invariant hierarchical models to analyse trial data from 117 meta-analyses in *ROBES* simultaneously. The models have been proposed in an earlier paper [9], but we describe the models briefly here to show how to derive the formulae for heterogeneity variance $\tau^2_{total,m}$ among all trials in a meta-analysis *m*.

### S1.1 Univariable model for the influence of accounting for a single trial design characteristic

In a given meta-analysis *m*, trials are categorised as low risk of bias (L-trials) or high/unclear risk of bias (H-trials) for a specific design characteristic.

The L-trials provide an estimate of the underlying intervention effect $\theta^L_{im}$, assumed to have a normal random-effects distribution with mean $d_m$ and variance $\tau^2_m$, specific to meta-analysis *m*. The H-trials are assumed to estimate an underlying intervention effect $\theta^H_{im}$, assumed to be normally distributed with mean $d_m + b_m$ and variance $\lambda \tau^2_m$:

$$\theta^L_{im} \sim N(d_m, \tau^2_m)$$
$$\theta^H_{im} \sim N(d_m + b_m, \lambda \tau^2_m).$$

The average bias $b_m$ in intervention effect in meta-analysis *m* is assumed to be exchangeable across meta-analyses, with overall mean $b_0$ and between-meta-analysis variance in mean bias $\varphi^2$:

$$b_m \sim N(b_0, \varphi^2)$$
$$b_0 \sim N(B_0, V_0)$$

We set an indicator $X_{im}$ to be 1 for H trials and 0 for L trials such that

$$X_{im} = \begin{cases} 1 & \pi_m \\ & \text{with probability} \\ 0 & 1 - \pi_m. \end{cases}$$

Each trial is assumed to provide an underlying estimate of intervention effect:

$$\theta_{im} = (1 - X_{im})\theta^L_{im} + X_{im}\theta^H_{im}.$$

The first term of the sum will return $\theta^L_{im}$ if trial *i* is at low risk of bias. The second term will return $\theta^H_{im}$ if the trial *i* is at high/unclear risk of bias.

The total heterogeneity variance among trials in meta-analysis *m* is given by:

$$\tau^2_{total,m} = \text{var}(\theta_{im})$$
$$= \text{var}((1-X_{im})\theta^L_{im} + X_{im}\theta^H_{im})$$
$$= \text{var}((1-X_{im})\theta^L_{im}) + \text{var}(X_{im}\theta^H_{im}) + 2\text{cov}((1-X_{im})\theta^L_{im}, X_{im}\theta^H_{im})$$

$$= E(1-X_{im})^2 \text{var}(\theta^L_{im}) + E(\theta^L_{im})^2 \text{var}(1-X_{im}) + \text{var}(1-X_{im})\text{var}(\theta^L_{im})$$
$$+ E(X_{im})^2 \text{var}(\theta^H_{im}) + E(\theta^H_{im})^2 \text{var}(X_{im}) + \text{var}(X_{im})\text{var}(\theta^H_{im})$$
$$+ 2[E((1-X_{im})\theta^L_{im} X_{im}\theta^H_{im}) - E((1-X_{im})\theta^L_{im})E(X_{im}\theta^H_{im})]$$

$$= E(1-X_{im})^2 \text{var}(\theta^L_{im}) + E(\theta^L_{im})^2 \text{var}(1-X_{im}) + \text{var}(1-X_{im})\text{var}(\theta^L_{im})$$
$$+ E(X_{im})^2 \text{var}(\theta^H_{im}) + E(\theta^H_{im})^2 \text{var}(X_{im}) + \text{var}(X_{im})\text{var}(\theta^H_{im})$$
$$+ 2[E((1-X_{im})X_{im})E(\theta^L_{im}\theta^H_{im}) - E((1-X_{im})\theta^L_{im})E(X_{im}\theta^H_{im})]$$

$$= (1-\pi_m)^2 \tau^2_m + d^2_m (1-\pi_m)\pi_m + (1-\pi_m)\pi_m \tau^2_m$$
$$+ \pi^2_m \lambda \tau^2_m + (d_m + b_m)^2 \pi_m(1-\pi_m) + \pi_m(1-\pi_m)\lambda \tau^2_m$$
$$- 2(1-\pi_m)d_m \pi_m(d_m + b_m)$$

$$= (1-\pi_m)\tau^2_m + d^2_m(1-\pi_m)\pi_m$$
$$+ \pi_m \lambda \tau^2_m + (d_m + b_m)^2 \pi_m(1-\pi_m)$$
$$- 2(1-\pi_m)d_m \pi_m(d_m + b_m)$$

$$= (1-\pi_m)\tau^2_m + \pi_m \lambda \tau^2_m + \pi_m(1-\pi_m)b^2_m$$

### S1.2 Multivariable model for the influence of accounting for multiple trial design characteristics

Suppose trials in a meta-analysis $m$ are categorised as low risk of bias (L-trials) or high/unclear risk of bias (H-trials) for each of 2 reported design characteristics. We set the indicator $X_{ijm}$ to be 1 for trials at high/unclear risk of bias for the $j$-th reported characteristic ($j=1,2$), and 0 for trials at low risk of bias for that characteristic such that

$$X_{ijm} = \begin{cases} 1 & \pi_{jm} \\ & \text{with probability} \\ 0 & 1-\pi_{jm} \end{cases}$$

Each trial is assumed to provide an estimate of underlying intervention effect:

$$\theta_{im} = (1-X_{1im})(1-X_{2im})\theta_{im}^L + X_{1im}(1-X_{2im})\theta_{1im}^H + X_{2im}(1-X_{1im})\theta_{2im}^H + X_{1im}X_{2im}\theta_{3im}^H$$

where

$$\theta_{im}^L \sim N(d_m, \tau_m^2)$$
$$\theta_{1im}^H \sim N(d_m + b_{1m}, \lambda_1\tau_m^2)$$
$$\theta_{2im}^H \sim N(d_m + b_{2m}, \lambda_2\tau_m^2)$$
$$\theta_{3im}^H \sim N(d_m + b_{1m} + b_{2m} + b_{3m}, \lambda_1\lambda_2\lambda_3\tau_m^2).$$

Trials at low risk of bias for both characteristics 1 and 2 provide an estimate of intervention effect $\theta_{im}^L$, as in Section S1.1. The intervention effect $\theta_{1im}^H$ in a trial $i$ at high/unclear risk of bias for characteristic 1 but low risk of bias for characteristic 2 has a normal distribution with mean $d_m+b_{1m}$ and variance $\tau_m^2\lambda_1$. The intervention effect $\theta_{2im}^H$ in a trial $i$ at high/unclear risk of bias for characteristic 2 but low risk of bias for characteristic 1 has a normal distribution with mean $d_m+b_{2m}$ and variance $\tau_m^2\lambda_2$. The intervention effect $\theta_{3im}^H$ in a trial $i$ at high/unclear risk of bias for both characteristics 1 and 2 has a normal distribution with mean $d_m+b_{1m}+b_{2m}+b_{3m}$ and variance $\tau_m^2\lambda_1\lambda_2\lambda_3$.

An estimate of total heterogeneity variance among trials in meta-analysis $m$ is given by:

$$\tau^2_{total,m} = \text{var}(\theta_{im})$$
$$= \text{var}((1-X_{1im})(1-X_{2im})\theta^L_{im} + X_{1im}(1-X_{2im})\theta^H_{1im} + X_{2im}(1-X_{1im})\theta^H_{2im} + X_{1im}X_{2im}\theta^H_{3im})$$
$$= \text{var}((1-X_{1im})(1-X_{2im})\theta^L_{im}) + \text{var}(X_{1im}(1-X_{2im})\theta^H_{1im}) + \text{var}(X_{2im}(1-X_{1im})\theta^H_{2im}) + \text{var}(X_{1im}X_{2im}\theta^H_{3im})$$
$$+ 2\text{cov}((1-X_{1im})(1-X_{2im})\theta^L_{im}, X_{1im}(1-X_{2im})\theta^H_{1im}) + 2\text{cov}((1-X_{1im})(1-X_{2im})\theta^L_{im}, X_{2im}(1-X_{1im})\theta^H_{2im})$$
$$+ 2\text{cov}((1-X_{1im})(1-X_{2im})\theta^L_{im}, X_{1im}X_{2im}\theta^H_{3im}) + 2\text{cov}(X_{1im}(1-X_{2im})\theta^H_{1im}, X_{2im}(1-X_{1im})\theta^H_{2im})$$
$$+ 2\text{cov}(X_{1im}(1-X_{2im})\theta^H_{1im}, X_{1im}X_{2im}\theta^H_{3im}) + 2\text{cov}(X_{2im}(1-X_{1im})\theta^H_{2im}, X_{1im}X_{2im}\theta^H_{3im})$$

$$= (1-\pi_{1m})(1-\pi_{2m})\tau^2_m + ((1-\pi_{1m})(1-\pi_{2m}) - (1-\pi_{1m})^2(1-\pi_{2m})^2)d^2_m$$
$$+ \pi_{1m}(1-\pi_{2m})\lambda_1\tau^2_m + (\pi_{1m}(1-\pi_{2m}) - \pi_{1m}^2(1-\pi_{2m})^2)(d_m + b_{1m})^2$$
$$+ \pi_{2m}(1-\pi_{1m})\lambda_2\tau^2_m + (\pi_{2m}(1-\pi_{1m}) - \pi_{2m}^2(1-\pi_{1m})^2)(d_m + b_{2m})^2$$
$$+ \pi_{1m}\pi_{2m}\lambda_1\lambda_2\lambda_3\tau^2_m + (\pi_{1m}\pi_{2m} - \pi_{1m}^2\pi_{2m}^2)(d_m + b_{1m} + b_{2m} + b_{3m})^2$$
$$- 2\pi_{1m}(1-\pi_{1m})(1-\pi_{2m})^2 d_m(d_m + b_{1m})$$
$$- 2\pi_{2m}(1-\pi_{2m})(1-\pi_{1m})^2 d_m(d_m + b_{2m})$$
$$- 2\pi_{1m}\pi_{2m}(1-\pi_{1m})(1-\pi_{2m})d_m(d_m + b_{1m} + b_{2m} + b_{3m})$$
$$- 2\pi_{1m}(1-\pi_{2m})\pi_{2m}(1-\pi_{1m})(d_m + b_{1m})(d_m + b_{2m})$$
$$- 2\pi_{1m}^2(1-\pi_{2m})\pi_{2m}(d_m + b_{1m})(d_m + b_{1m} + b_{2m} + b_{3m})$$
$$- 2\pi_{2m}^2(1-\pi_{1m})\pi_{1m}(d_m + b_{2m})(d_m + b_{1m} + b_{2m} + b_{3m}).$$

In a similar way, we derive estimates of total heterogeneity in a meta-analysis from the multivariable label-invariant models for the influence of accounting for three design characteristics.

**S2 Model comparison**

Bayesian hierarchical models were fitted to trial data from all 117 meta-analyses. The various models fitted to the data differed according to the indicators of design characteristics and interactions included as covariates in the model, and according to the inclusion of indicators of outcome type in the regression model for heterogeneity variance $\tau_m^2$ among trials at low risk of bias. Results to compare model fit are given in Table S1.

**Table S1** Posterior mean residual deviance $D_{res}$, effective number of parameters $p_D$ and deviance information criterion (DIC) for the hierarchical models fitted to the ROBES data.

| Model | Design characteristic/s | Interaction/s between design characteristics | Covariates in model for $\tau^2$ | $D_{res}$ | $p_D$ | DIC |
|---|---|---|---|---|---|---|
| A1 | Sequence generation | N/A | - | 2982 | 1909 | 4891 |
|    | Sequence generation | N/A | Outcome type | 3000 | 1889 | 4889 |
| A2 | Allocation concealment | N/A | - | 2972 | 1914 | 4886 |
|    | Allocation concealment | N/A | Outcome type | 3003 | 1899 | 4902 |
| A3 | Blinding | N/A | - | 2968 | 1915 | 4883 |
|    | Blinding | N/A | Outcome type | 3003 | 1891 | 4894 |
| B1 | Sequence generation and allocation concealment | Yes | - | 3001 | 1900 | 4901 |
|    | Sequence generation and allocation concealment | No | - | 2988 | 1906 | 4894 |
| B2 | Sequence generation and blinding | Yes | - | 2978 | 1908 | 4886 |
|    | Sequence generation and blinding | No | - | 2988 | 1904 | 4892 |
| B3 | Allocation concealment and blinding | Yes | - | 2996 | 1902 | 4898 |
|    | Allocation concealment and blinding | No | - | 2981 | 1908 | 4889 |
| B4 | Sequence generation, allocation concealment and blinding | All possible | - | 2985 | 1905 | 4890 |
|    | Sequence generation, allocation concealment and blinding | Interaction between sequence generation and blinding alone | - | 2998 | 1899 | 4897 |
|    | Sequence generation, allocation concealment and blinding | No | - | 2978 | 1913 | 4891 |
|    | Sequence generation, allocation concealment and blinding | No | Outcome type | 2991 | 1895 | 4886 |